\documentclass{emulateapj}

\shorttitle{Magneticallly-dominated outflow in GRB 080916C} 
\shortauthors{Zhang \& Pe'er}

\newcommand{\GeV}{\rm{\, GeV }}

\newcommand{\beq}{\begin{equation}}
\newcommand{\eeq}{\end{equation}}
\newcommand{\ba}{\begin{array}}
\newcommand{\ea}{\end{array}}

\newcommand{\ee}{\epsilon_{e,0}}

\def \etal{{\it et al.~}}
\def\be{\begin{equation}}
\def\ee{\end{equation}}
 
 \def\gtrsim{\raisebox{-0.3ex}{\mbox{$\stackrel{>}{_\sim} \,$}}}

\begin{document}
\title{Evidence of an initially magnetically dominated outflow in GRB 080916C}

\author{Bing Zhang\altaffilmark{1}}
\author{Asaf Pe'er\altaffilmark{2}\altaffilmark{3}}

\altaffiltext{1}{Department of Physics and Astronomy,
 University of Nevada, Las Vegas, NV 89154, USA}
\altaffiltext{2}{Space Telescope Science Institute, 
Baltimore, MD 21218, USA}
\altaffiltext{3}{Riccardo Giacconi Fellow}

\begin{abstract}
  The composition of gamma-ray burst (GRB) ejecta is still a
  mystery.  The standard model invokes an initially hot ``fireball''
  composed of baryonic matter.
  Here we analyze the broad band spectra of GRB 080916C
  detected by the Fermi satellite. 
  The featureless Band-spectrum of all five epochs as well as 
  the detections of 
  $\gtrsim 10 $~GeV photons in this burst place a strong constraint
  on the prompt emission radius $R_\gamma$, which is typically
  $\gtrsim 10^{15}$ cm, 
  independent on the details of the emission process. 
  The lack of detection of a thermal component as 
  predicted by the baryonic models
  strongly suggests that a 
  significant fraction of the outflow energy is initially
  not in the ``fireball'' form, but is likely in a Poynting
  flux entrained with the baryonic matter. The ratio between
  the Poynting and baryonic fluxes is at least $\sim (15-20)$
  at the photosphere radius, if the Poynting flux is not
  directly converted to kinetic energy below the photosphere.
\end{abstract}


\slugcomment{2009, ApJ, 700, L65-L68}

\section{Introduction}
\label{sec:intro}

In the classical model of Gamma-ray bursts (GRBs), an
initially hot ``fireball'' is composed of photons, electron/positron
pairs, and a small amount of baryons
\citep{paczynski86,goodman86,shemi90}. This fireball is soon
accelerated to a relativistic speed under its own thermal pressure.
Due to the existence of baryons, a significant fraction of energy is
converted into the kinetic energy of the ejecta \citep{meszaros93,
  PSN93}. The rest of energy is still stored in the form of photons,
which escape the system as the fireball becomes transparent at the
photosphere radius \citep{meszaros00,MRRZ02}. The ejecta then coast
with a relativistic speed without significant radiation until reaching
a larger radius when a fraction of kinetic energy is 
dissipated into heat and radiation in internal shocks \citep{rees94}.
The internal shock model has the advantage of interpreting GRB
variabilities, but suffers the low radiation efficiency
problem \citep{Kumar99, PSM99}. Recently, an analysis of the prompt 
emission data of several GRBs suggests
that the internal shock model is disfavored by the data
\citep{kumar08,kumar09}.
An alternative model invokes a dynamically important
magnetic field. Within such a Poynting flux dominated model, the
observed GRB emission is powered by dissipation of the magnetic energy
within the ejecta \citep{usov94,Thompson94,vlahakis03,MR97b,lyutikov03}. 

Until recently, it has been difficult to diagnose the GRB 
composition from observational data.
The recent detection of the broad-band featureless Band-function
spectra as well as the
very high gamma-ray emission $(\gtrsim 10 \GeV$) from GRB 080916C by
the Fermi satellite \citep{abdo09} provides a unique opportunity
to diagnose the GRB composition\footnote{An 18-GeV photon was detected
earlier from GRB 941017 \citep{hurley94}. However, it was
significantly delayed, not associated with the prompt
emission. The bandpass of prompt emission observation was not wide
enough to constrain the thermal emission component.  Hence, it
could not be applied to directly constrain
the GRB emission radius and composition.}.
Below we will use the opacity argument (\S2) and the photosphere
argument (\S3) to argue that the ejecta of GRB 080916C 
must contain a significant fraction of energy
that is initially not in the ``fireball'' form, but in a Poynting
flux entrained with baryonic matter. 

\section{Model independent emission radius constraint}
\label{sec:2}

Both the Large Area Telescope (LAT) and the Gamma-ray Burst Monitor
(GBM) on-board Fermi have detected GRB 080916C.  The time-dependent
spectral analysis reveals a series of featureless smoothly-joint
broken power law \citep[``Band-function'',][]{band93} 
spectrum categorized by a peak energy
$E_p$ and two asymptotic photon spectral power law indices $\alpha$
and $\beta$ \citep{abdo09}.  These spectra cover 5-6 orders of
magnitude in energy, from $\sim 10$ keV to $\sim (1-10)$ GeV.  The
highest photon energy reaches 13.2 GeV (in the time interval ``d''
defined in Fig.1 of \cite{abdo09}).  At a redshift $z=4.35\pm 0.15$
\citep{greiner09}, this burst is the most energetic GRB known to date,
with an isotropic gamma-ray energy $\sim 9\times 10^{54}$ erg.

The gamma-ray spectrum is expected to have a pair cutoff feature at
large energies due to the compactness argument, i.e. the optical depth
for two photon pair production ($\gamma\gamma \rightarrow e^{+}e^{-}$)
may reach unity above a critical cutoff energy $E_{cut}$.  Within the
internal shock model, the pair cutoff energy (or the lack of it),
together with the observed variability time scale, can be used to
constrain the bulk Lorentz factor $\Gamma$ of the outflow
\citep{baring97,lithwick01,abdo09}. In some models, the observed
variability time scale may not reflect that of the central engine
\citep{lyutikov03,narayan09}. So more generally the pair cutoff energy
can be expressed as a function of two independent parameters, the bulk
Lorentz factor $\Gamma$ and the gamma-ray emission radius $R_\gamma$
from the central engine \citep{gupta08}. For each cutoff energy
$E_{cut}$, one can define a threshold energy $E_{td}$ above which the
photons with this energy can interact with the photons at $E_{cut}$ to
produce pairs. This threshold condition is defined by $(E_{td}/{1~{\rm
MeV}})(E_{cut}/{1~{\rm MeV}}) \gtrsim 0.25 [\Gamma/(1+z)]^2$.  The
expression of the pair production optical depth $\tau_{\gamma\gamma}$
depends on the relative location between $E_{td}$ and the
Band-function break energy. For $E_p(\alpha-\beta)/(2+\alpha) < E_{td}
< E_{cut}$ (which is generally satisfied for this analysis), the
optical depth of a photon with the observed energy $E$ can be coasted
into a simple form (derived from Eqs.(13) and (14) of Gupta \&
Zhang 2008, or from Eqs. (3-4) of Lithwick \& Sari 2001 with
$\delta t$ absorbed into the expression of $R_\gamma$ and
with the cosmological correction factor $(1+z)$ properly taken
into account)
\begin{equation}
\tau_{\gamma\gamma}(E)=\frac{C(\beta) \sigma_T d_z^2 f_0}{-1-\beta}
\left(\frac{E}{m_e^2c^4}\right)^{-1-\beta} \frac{1}{R_\gamma^2}
\left(\frac{\Gamma}{1+z}\right)^{2+2\beta}~,
\label{tau}
\end{equation}
where $d_z=(c/H_0)\int_0^z dx/\sqrt{\Omega_\Lambda+\Omega_m
(1+x)^3}$ is the comoving distance of the GRB, $z$ is the redshift
of the GRB, and $m_e$, $c$ and $\sigma_T$ are the fundamental
constants electron mass, speed of light, and Thomson cross section,
respectively. Here the energies, i.e. $E$ and $m_e c^2$, are both
in units of ``keV''. The parameter $f_0$ (in units of 
${\rm ph \cdot cm^{-2} \cdot (keV)^{-1-\beta}}$)
is such defined that the observed photon 
{\em fluence} spectrum above the break is written as 
${\cal N}(E) = f_0 E^{\beta}$.
In terms of the fitting parameters of \cite{abdo09},
$f_0$ can be expressed as
\begin{equation}
f_0=A \cdot \Delta T\left[\frac{E_p(\alpha-\beta)}{(2+\alpha)}
\right]^{\alpha-\beta} \exp(\beta-\alpha)
(100~{\rm keV})^{-\alpha}~,
\end{equation}
where $\Delta T$ is the (observed) time interval taken to perform the
Band-function fit, and $A$ is the Band-function normalization
in units of ${\rm ph \cdot cm^{-2} \cdot s^{-1} \cdot {keV}^{-1}}$.
Finally, the coefficient $C(\beta)$ in Eq.(1) is a function of 
$\beta$, and we adopt the approximation $C(\beta) \simeq (7/6)
(-\beta)^{-5/3}/(1-\beta)$ \citep{svensson87} to perform
the calculations\footnote{\cite{gupta08} takes
$C(\beta)=3/8(1-\beta)$, while \citep{lithwick01} takes 
$C(\beta)=11/180$. For a typical value $\beta=-2$, the 
$C(\beta)$ value of \cite{svensson87} and \cite{gupta08}
are consistent with each other, but that of \cite{lithwick01}
is smaller by a factor of $\sim 2$.
}.

The lack of a spectral cutoff feature suggests
$\tau_{\gamma\gamma}(E_{\rm max}) \leq 1$, where $E_{\rm max}$ is the
maximum energy of the observed photons.  Using this condition, one can
derive the $\Gamma-R_\gamma$ constraints for the 5 time intervals
(a-e) defined by \cite{abdo09} (using the values in their Table 1,
with $\Delta T$ defined by the time ranges listed in the table).
By doing so, we have implicitly assumed that the emission location
and Lorentz factor for a particular time interval $\Delta T$ 
essentially remain the same. 
Figure 1 shows the five critical lines (solid) in the
$\Gamma-R_\gamma$ space. The allowed parameter regimes are above the
lines. For all the 5 time intervals the allowed emission radii are all
large. This model-independent conclusion regarding a large $R_\gamma$
is consistent with the results derived from other independent methods
for other GRBs \citep{lyutikov06,kumar08,racusin08,kumar09,shen09}.

\begin{figure}
\plotone{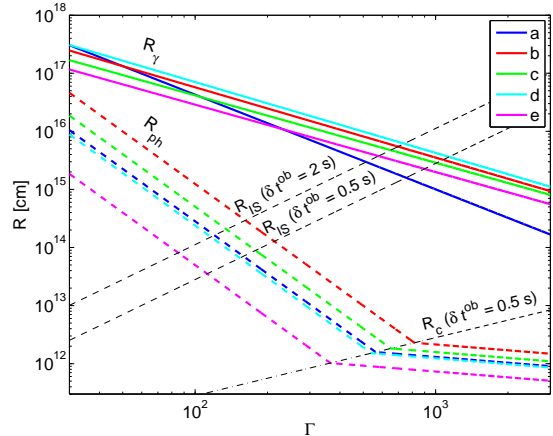}
\caption{\protect
The $\Gamma-R$ diagram of GRB 080916C. The constraints on $R_\gamma$
are displayed in color thick lines, above which are the allowed 
parameter spaces for each corresponding epoch. 
The photosphere radii $R_{ph}$ are displayed as color dashed lines,
with the same color convention. The internal
shock radii $R_{\rm IS}=\Gamma^2 c \delta t^{ob}/(1+z)$ for $\delta t^{ob}
=0.5, 2$ s are the black dashed lines.
}
\label{figure:1}
\end{figure}

Swift observations reveal a commonly seen steep-decay phase in the
early X-ray afterglow \citep{tagliaferri05}, which suggests that the
GRB prompt emission region is separated from the emission region of
the afterglow (the external shock) \citep{zhang06}. Within 
the ``internal'' models, the expected dissipation radius spans in
a wide range, from the photosphere radius (typically at $R_{ph}
\sim (10^{11}-10^{12})$ cm) to slightly smaller than the deceleration
radius ($\sim 10^{17}$ cm).
The derived large emission radius $R_\gamma$ is inconsistent with the
photosphere model (see \S3 for detail), but is consistent with
that expected from the magnetic dissipation model \citep{lyutikov03}.  
The ``internal shock'' model invokes
an emission radius $R_{\rm IS} \sim \Gamma^2 c \delta t$ (where
$\delta t$ is the variability time scale of the central engine). The
$\Gamma- R_{\gamma}$ constraint of this model is plotted in
figure \ref{figure:1} as black dashed lines 
for two values of the variability time scales
discussed in \citet{abdo09}, $\delta t = 0.5, 2$~s. The internal shock
site is allowed by the data. The constrained internal shock radii are
at least $R_{\rm IS} \sim 10^{15}-10^{16}$ cm for the five epochs. The
required Lorentz factors are at least 500-1000 (see Fig.1).  However,
there are further constrains on this model, as discussed below.

\section{Constraints from the photospheric  (thermal) emission}
\label{sec:3}

Besides the energy dissipation region (internal shock or magnetic
dissipation site), the fireball ``photosphere'',
at which the fireball becomes transparent during the expansion,
is another emission site of GRBs
\citep{meszaros00,rees05,thompson07,ghisellini07,lazzati09}.
The photosphere radius $R_{ph}$ is defined by the electron scattering
optical depth $\tau'_{\gamma e} = n'\sigma_T \Delta'=1$, where $n'$
and $\Delta'$ are the electron number density and width of the ejecta
shell in the rest frame comoving with the ejecta. Below we will 
assume a pure baryonic flow (as expected in the fireball model) to
derive the expected photosphere spectra. For a continuous
baryonic wind (which is suitable to describe GRB 080916C given the
smooth lightcurves reported in Fig.1 of \cite{abdo09}) with a total
wind luminosity $L_w$, the photosphere radius can be written as
\citep{meszaros00,Peer08}
$R_{ph} \simeq ({ L_w\sigma_T R_0^2 / 8\pi m_p c^3 \eta})^{1/3}$ 
for $R_{ph} < R_c$, and
$R_{ph} \simeq { L_w\sigma_T /  8\pi m_p c^3 \Gamma^2 \eta}$  
for $R_{ph} > R_c$.
Here, $m_p$ is the proton mass, $\eta=L_w/\dot Mc^2$
is the dimensionless entropy of the baryonic flow, and 
$R_c \sim R_0 \cdot {\rm min}(\eta,\eta_*)$ is the radius at which 
the ejecta wind reaches the ``coasting'' phase. The critical 
dimensionless entropy is $\eta_* = (L_w\sigma_T/8\pi
m_p c^3 R_0)^{1/4}$, and $R_0 = c \delta t$ is the radius where the
ejecta is emitted from the central engine.  The coasting Lorentz
factor is $\Gamma=\eta$ for $R_{ph} > R_c$, and $\Gamma=\eta_*$ for
$R_{ph} \leq R_c$. 
The photosphere radii for the 5 epochs of GRB 080916C
are presented in Fig.1 by the color dashed lines.

It has been suggested that the observed $E_p$ of GRBs may be the
thermal peak of the photosphere emission 
\citep{rees05,thompson07},
and that the Comptonization of the thermal emission may lead to a
non-thermal spectrum above $E_p$ \citep{PMR05, peer06}. However, for
GRB 080916C, Figure 1 suggests that the observed gamma-ray emission is
{\em not} from the photosphere, since $R_\gamma \gg R_{ph}$. Under the
high-compactness condition, a second pair photosphere may be formed
\citep{MRRZ02,kobayashi02,peer04}, but its radius 
is also much smaller than $R_\gamma$
inferred from the data. Another suggestion is
that the observed GRB spectra are the superposition of a thermal
component (photosphere) that defines $E_p$ and a non-thermal power law
component \citep{ryde05,ryde08}. This model is based on the
data in a relatively narrow spectral range of {\it BATSE}, and
when extrapolated to high energies, would violate the observed
featureless Band-function significantly. 
The theoretically expected thermal peak energy $\approx 50$~keV (see
below) is well below the observed $E_p$ in the 5 epochs discussed by
\citet{abdo09}. Clearly this model by itself is insufficient
to interpret the data of this burst.

The initial wind luminosity $L_w$ of the fireball is at least the
observed gamma-ray luminosity $L_\gamma$ (assuming radiation
efficiency 100\%), i.e. $L_w \geq L_\gamma$.  Such an outflow, during
the expansion, must have released a residual thermal emission at the
transparent (photosphere) radius. The luminosity of this thermal
component is high, and can be written as \citep{meszaros00}
\begin{equation}
L_{th}= \left\{
\begin{array}{ll}
L_w, & \eta>\eta_*,~ R_{ph} < R_c, \\
L_w (\eta/\eta_*)^{8/3}, & \eta<\eta_*,~ R_{ph} > R_c.\\
\end{array}
\right.
\end{equation}
The expected temperature of the black body component emerging from the
photosphere is thus 
$T_{ph}^{ob}= (L_w / 4\pi R_0^2 c a)^{1/4} (1+z)^{-1}$ 
for $R_{ph} < R_c$, and $T_{ph}^{ob}=(L_w / 4\pi R_0^2 c a)^{1/4}
(R_{ph}/R_c)^{-2/3} (1+z)^{-1}$ for $R_{ph} > R_c$, respectively,
where $a$ is Stefan-Boltzmann's constant.

Assuming a blackbody spectrum (other modified spectral shapes would 
not change the conclusion), we plot in Fig.2 the {\em lower limit} of
the expected photosphere spectrum for the internal shock model in
the baryon-dominated outflow (by taking $L_w=L_\gamma$) and 
compare it with the observations. For clarity, only the prediction
for the epoch (b) are plotted (red thick-dashed curve). This plot 
corresponds to $\eta\geq \eta_* = 850$, hence $L_{th} = L_w$, 
$R_{ph} \simeq R_c$ and $T_{ph}^{ob}= T_{ph,\max}^{ob} = 50$~keV. 
It is apparent that such a component
strongly violates the observational data, which is a featureless
Band-function. 

One can further argue that the baryonic model in general does
not work. In Fig. 2, we present the predicted lower limits of the
photosphere spectrum for two additional (more conservative) 
temperatures $T_{ph}^{ob} = 10, 1$~keV.
These correspond to either a coasting Lorentz factor $\eta = 470,
200$ (for the same $\delta t^{ob}$), or a central engine variability 
time scale much longer than what is observed (for the same $\eta$),
which may be relevant to a fireball emerging from a
stellar envelope in the collapsar model. These models give a
much larger $R_{ph}$, and hence, a much lower $T_{ph}^{ob}$.
Since $\delta t^{ob}=0.5$ s
has been observed \citep{abdo09}, the models that invoke a longer
central engine variability time scale require the unconventional
assumption that the observed variability time scale is not that of the 
central engine \citep{narayan09}. Nonetheless, the predicted 
thermal spectra of these cases (red, thin-dashed curves) also 
strongly violate the observational constraints.

This analysis strongly suggests that the initial wind luminosity
was not stored in the ``fireball'' form at the base of the central
engine. Since there is no other known source of energy that can
be entrained in the ejecta, we identify the missing luminosity as the
Poynting flux luminosity, which is not observable before strong
magnetic dissipation occurs at a much larger radius. It has been
suggested \citep{zhang02,daigne02} that the photosphere thermal
component can be much dimmer if the outflow is Poynting flux
dominated. In order to hide the bright thermal component, one can 
pose a lower limit on the 
ratio between the Poynting flux and the baryonic flux,
$\sigma \equiv L_P/L_b$,. Following
\cite{zhang02}, we define $L_w=L_b+L_P=(1+\sigma)L_b$. Assuming
no dissipation of the Poynting flux below $R_{ph}$, the above
photosphere derivations can be modified by replacing $L_w$
by $L_w/(1+\sigma)$. One can then derive the required minimum
$\sigma$ value that can ``hide'' the photosphere thermal
component.  For the specific internal shock
model required by the variability time scale (the red thick-dashed
curve), one requires at least $\sigma \simeq 20$ at the
photosphere to make the photosphere
emission unobservable (red thick-dotted curve in Fig.2).  
At such a high-$\sigma$ shocks are very weak and cannot power the 
bright gamma-ray emission \citep{zhang05},
so unless $\sigma$ can reduce significantly from the photosphere
to the internal shock radius (e.g. Spruit et al. 2001) 
but without generating a dominant emission component in the observed 
spectrum, the internal shock model is not a viable
mechanism to interpret this burst\footnote{Another possibility
would be that the Poynting flux is converted to kinetic energy
directly without magnetic dissipation during the acceleration 
phase \citep[e.g.][]{vlahakis03}. The photosphere brightness of 
this model is not studied in detail.}.
For the other two unconventional cases, similar minumim
values of $\sigma = 15 - 20$ at photosphere 
are required in order to obtain
consistency with the observed spectra. We can therefore conclude 
that the ejecta of GRB 080916C cannot be a pure 
``fireball'', but must store a significant fraction
of energy initially in a Poynting flux.

\begin{figure}
\plotone{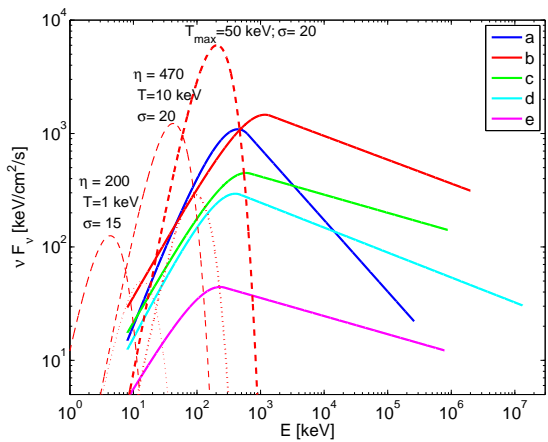}
\caption{\protect
The observed Band-function spectra for the five epochs \citep{abdo09}
(color solid) and the predicted lower limits of the photosphere
spectra (red dashed) for 
different parameters for the epoch (b) within the framework of
the baryonic fireball models. Red, thick-dashed curve: the internal
shock model with $\delta t^{ob} =0.5$ s, corresponding to 
$T_{ph}^{ob}=50$ keV;
red, thin-dashed curves: for $T_{ph}^{ob}=10, 1$ keV. The suppressed 
photosphere spectra are plotted by red, dotted curves, with the
required $\sigma$ values marked.
}
\label{figure:2}
\end{figure}

\section{Summary and discussion}
\label{sec:summary}

In this {\it letter}, we show that the observed feature-less
Band-function broad band spectra of GRB
080916C observed by the {\it Fermi} satellite \citep{abdo09}
pose two interesting constraints on the GRB prompt emission mechanism.
First, the detection of high energy, $\gtrsim 10$~GeV,
photons places a strong model-independent $\Gamma- R_{\gamma}$ 
constraint on the GRB ejecta (Fig.1),
which precludes the photosphere ($R_{ph} \ll R_{\gamma}$) 
as the emission site. Second, the non-detection of a bright 
thermal component
expected from the baryonic models puts a
strong constraint on the
composition of the fireball: the flow should be
dominated by a Poynting flux component
with a minimum $\sigma \approx 15-20$ at the photosphere
in order to account for the observed spectra (Fig. 2),
as long as the Poynting flux energy is not directly converted 
to the kinetic energy of the flow below the photosphere.

Within such a picture, the observed bright gamma-ray
emission may be powered by dissipation of the Poynting flux energy
within the outflow. 
A Poynting flux
dominated flow favors synchrotron radiation as the emission
mechanism of the observed gamma-rays. This is consistent with
independent modeling of the burst \citep{wang09}. It also
suggests a weak synchrotron self-Compton component, which is not
observed from the data.
This model requires that the GRB central engine launches a
collimated magnetic bubble from the central engine (black hole or 
millisecond magnetar), which propagates through the stellar 
envelope without reducing the $\sigma$ value significantly
\citep{wheeler00}. 
After escaping the star, the ejecta is quickly 
accelerated under its own magnetic pressure \citep{mizuno09}. At the 
radius where photons become transparent (photosphere), the bulk of 
the wind energy is not in the thermal form. 
The $\sigma$-value
may reduce above photosphere. However, in order to have internal
shocks, the magnetic energy needs to be ``quietly'' dissipated 
below the internal shock radius without significant emission.
More naturally, the magnetic field may be globally dissipated
in a cataclysmic manner to power the observed emission.
This model does not demand a high $\sigma$ at the deceleration 
radius, since $\sigma$ of the flow would 
decrease significantly in the prompt 
emission region due to intense magnetic dissipation. In any
case, the $\sigma$ value at the deceleration radius can be of order
unity or higher, which gives a weak or moderately bright reverse
shock emission component \citep{zhang05,mimica09}.

This analysis applies to GRB 080916C. Similar analyses can be
carried out for more bursts co-detected by Fermi LAT/GBM in 
the future to determine whether other GRB outflows are also 
magnetically dominated.

\acknowledgments
We thank F. Ryde for discussion on the Fermi data, M. Baring, 
Z. G. Dai, Y. Z. Fan, G. Ghisellini, D. Giannios,
P. Kumar, E. P. Liang,
P. M\'esz\'aros, S. Nagataki, K.-I. Nishikawa, C. Thompson, 
K. Toma, N. Vlahakis, X. Y. Wang, X. F. Wu, and H. Yan for 
discussions/comments.
BZ acknowledges NASA NNG05GB67G, NNX08AN24G and NNX08AE57A,
and AP acknowledges the Riccardo Giacconi Fellowship at the
Space Telescope Science Institute for supports.

\end{document}